# The Emergence of Animal Social Complexity: theoretical and biobehavioral evidence

Bradly Alicea
Orthogonal Research (http://orthogonal-research.tumblr.com)

**Abstract**

This paper will introduce a theory of emergent animal social complexity using various results from computational models and empirical results. These results will be organized into a vertical model of social complexity. This will support the perspective that social complexity is in essence an emergent phenomenon while helping to answer two interrelated questions. The first of these involves how behavior is integrated at units of analysis larger than the individual organism. The second involves placing aggregate social events into the context of processes occurring within individual organisms over time (e.g. genomic and physiological processes). By using a complex systems perspective, five principles of social complexity can be identified. These principles suggest that lower-level mechanisms give rise to high-level mechanisms, ultimately resulting in metastable networks of social relations. These network structures then constrain lower-level phenomena ranging from transient, collective social groups to physiological regulatory mechanisms within individual organisms. In conclusion, the broader implications and drawbacks of applying the theory to a diversity of natural populations will be discussed.

**Introduction**

Animal social complexity, like many other multifaceted processes, is a complex system. As a complex system, extracting causal relationships from the multitude of interactions can be a daunting proposition. Drawing parallels with social media platforms (Churchill and Halverson, 2005), not only is animal social complexity rich with interactions, but is also centered around the concept of social networks (Pinter-Wollman et.al, 2013). Yet these parallels reveal that behavior alone is not enough to explain the dynamics of social organization over time. In this paper, I will present a theory that links the emergence and dynamics of social complexity to genomic and physiological mechanisms. Such a vertical approach reveals many opportunities for transient phenomena to emerge (e.g. collective behaviors, phenotypes linked to social status), which in turn serve as feedback and other regulatory mechanisms. This theory will also require support derived from complex systems principles, various behavioral models, and empirical examples, to which we will now turn.

**Principles**

From observations and theoretical insights into complexity across various types of complex systems (Allen and Starr, 1982, Kelso, 1995, Kauffman, 1993), five principles of animal social complexity can be identified (Figure 1). The first (Figure 1, I) suggests that systems are composed of multiple hierarchical layers, all of which interact with each other. This leads to the second principle (Figure 1, II), which is that interactions at each layer contribute to the formation of patterns at other levels of organization (Haken, 1977). While the self-organizing effects of these two principles are most striking and easily measured in physical and chemical systems, they also play a role in imposing order (third principle – Figure 1, III) upon behavioral and social systems (Batten, 2000).

Given this observation, a fourth principle (Figure 1, IV) can be proposed: the interplay between different lower-level components such as hormonal concentrations, neural structures, and genes can affect patterns of interactions at the organismal level. Lumdsen and Wilson (1980) have referred to this transformation of stimuli into thoughts and social bonds as "epigenetic" in that the products of both the lower-level and organismal level ultimately affects the social level. This leads to the fifth principle (Figure 1, V), which postulates that social groups exist as metastable networks (Batten *et al, 19*95). This results in certain key parameters of the system being attracted to a finite group of stable states over time. This may be representative of distinct species, distinct genotypes and phenotypes, or generally representative over time of transient social groups.

### Fundamental Units of Social Complexity

There are three units of analysis upon which the models and empirical work presented here will rely. The first unit type in is composed of individuals that possess an underlying physiology. The second unit type is called a dyad, and can be defined as two conspecifics engaged in an interaction. Jager and Segel (1992) define such encounters within a population of organisms as pairwise interactions, during which the role and status of each individual are defined. Models of cultural transmission and evolution (Boyd and Richerson, 1985; Cavalli-Sforza and Feldman, 1981) use this unit of analysis, and define an interaction as some observation or exchange of information or signals. Information can involve instructions about the world in general, while signals refer more to phenotypic markers and sensory stimuli. In this context, pairings usually involve organisms with differential degrees of overall dominance (Hemelrijk, 2000).

From a systems standpoint, Hogeweg and Hesper (1985), Theraulaz *et al* (1995), and Bonabeau *et al* (1996) argue that generally dominance orders in societies result from self-organizing processes driven by a double positive feedback mechanism. Given a single set of dyadic interactions over time, winners reinforce their probability of winning while losers reinforce their probability of losing (Hogeweg and Hesper, 1983). Over time, this can lead to patterns of interaction between both two organisms and between social units of many organisms apiece.

The third unit type is small groups, loosely defined by Arrow *et al* (2000) as three or more conspecifics engaged in an interaction. The small groups not only build on patterns that emerge from lower levels, but also act as units of selection in and of Tarpy *et al* (2004) use the idea of within-colony and between-colony selection to examine Queen replacement in honey bee colonies. When between-colony selective pressures predominate, cooperation is expected among the small groups. Different studies examining diverse behaviors in different taxa have used the terms band, flock, swarm, and colony to aid in delimiting and identifying complex patterns of social complexity. Now that three analytical units have been established, the principle of metastability will now be explored in more detail.

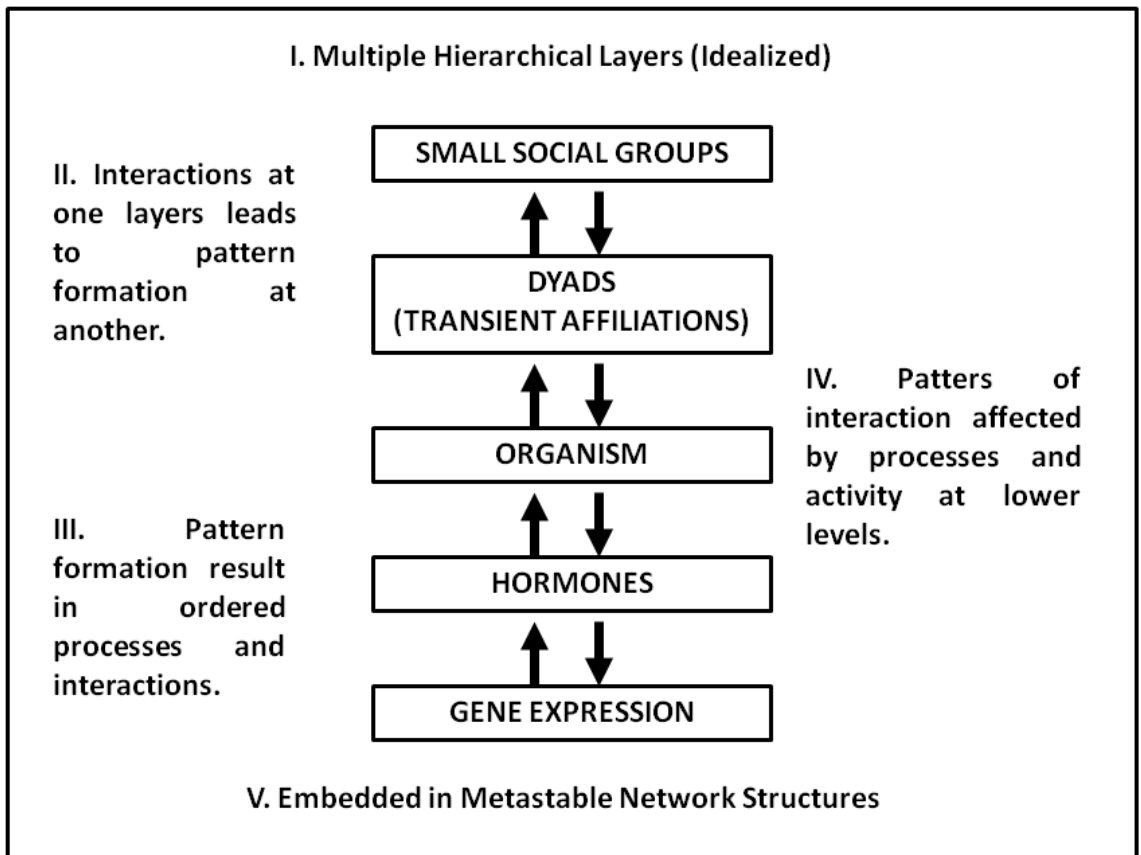

**Figure 1.** Five principles (I-V) of animal social complexity in context.

**Multistability and Social Complexity**

According to Kelso (1995), neural, behavioral, and ultimately social systems are multistable (Figure 2). Any set of quantitatively measured interactions ultimately form a function on a phase space, the trajectory of which is representative of the outcome for a time series as plugged into a set of differential equations. Read (1997) characterizes these order parameters as the outcome of intertroop interactions among nonhuman Primates. The *plateaus* and *valleys* in a phase space represent stable states, while the areas in-between represent systems in transition. In his theory of synergetics, Haken (1977) tells us that all systems originate at some initial state, and evolves over time. At certain critical periods, changes in the order parameters force the trajectory of a system to bifurcate into more than one stable state. As empirical examples can attest, such *pitchfork bifurcations* may result from the system reaching a threshold or encountering ambiguity.

The idea of multiple stable states is consistent with (but not identical to) the concept of dynamic equilibrium (Sole and Goodwin, 2001) and the more pervasive idea of homeostasis (Odum, 2000). While it is a subtle concept, the clustering behaviors in human societies may be understood as social networks favoring evolution towards metastability (Mas, Flache, and Helbing, 2010) Using a synergetics model also implies that complex social systems are perpetually at the boundary of a state transition. It can be argued that what makes emergent social complexity such a difficult concept to measure using reductionist methods is its propensity for sudden transitions and itinerant dynamics. Indeed the synergetic model works best when describing the transition of one physical state to another; as applied to human behavioral data it has worked best describing perceptual changes in relation to stimuli such as ambiguous figures (Kelso, 1995). Yet nonlinear, bistable dynamics can also be used to explain honeybee foraging dynamics (Loengarov and Tereshko, 2008).

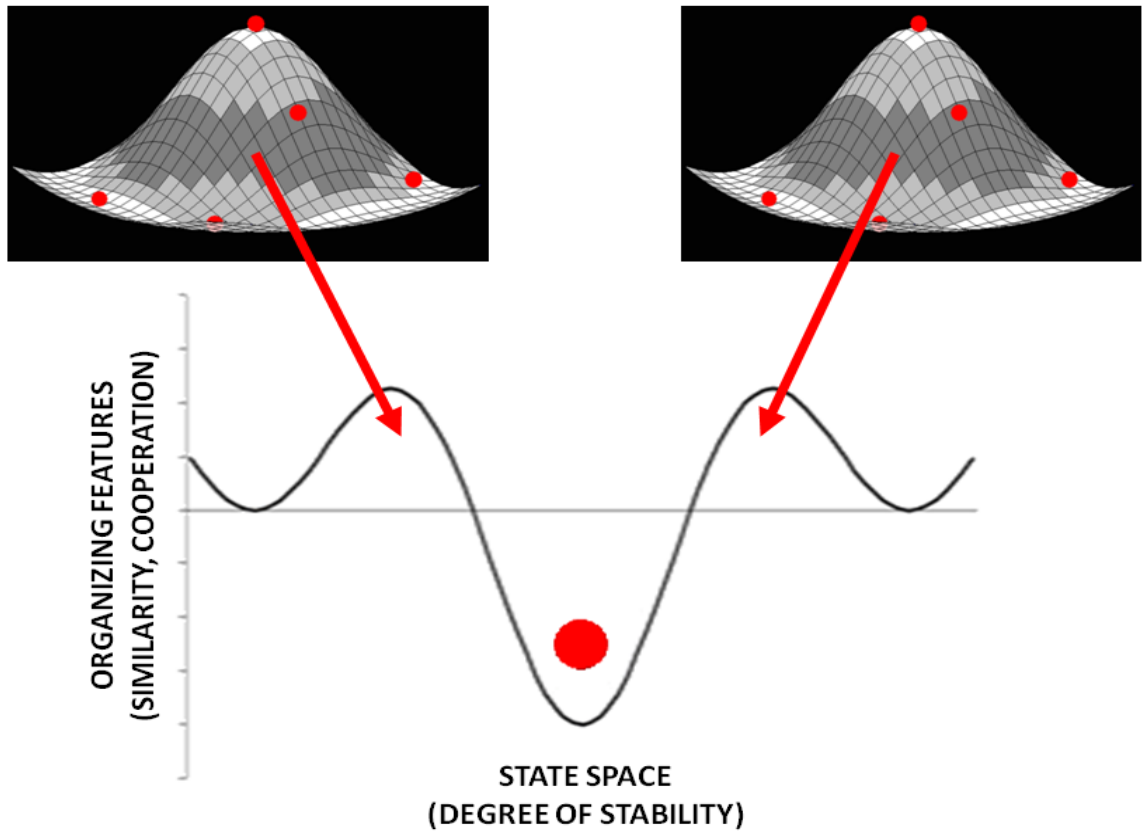

**Figure 2.** A schematic demonstrating the multistability concept (as proposed by synergetics).

Bonabeau (1997) and deVries and Biesmeijer (2002) explore the concepts of symmetry breaking and cross-inhibition among social insects in the context of multistable complex systems. deVries and Biesmeijer (2002) introduce symmetry breaking as the number of foragers visiting two equally profitable food sources will diverge after some time. The related concept of cross-inhibition is the phenomenon that, by increasing the profitability of one of two equal food sources, the number of foragers visiting the other source will decrease. This poses two fundamental questions that can be answered using these methods. The first of these relates to how organisms are recruited to each site, and the second involves how this contributes to formations of stable behavioral and social states. The concept of herd strategy sorting (Banerjee, 1992) among mammals will likewise rely on the pitchfork bifurcation as a model for guiding behavioral and social changes over time.

In general, interactions between individuals over space and time are driven by properties of the individual such as aggression, physiology, and memory (Chase *et al*, 1994) and observations of the environment. Kauffman (1993) provides evidence that game-theoretic stable states arise through the strategic interaction of multiple agents drive the system to one stable state or another. Newman (2003) also observes that in general, humans in social networks act upon information gathered from first-order connections only. This can lead to hierarchical differentiation within the local population (Hemelrijk, 2000) over time, which exists either in a chaotic or stable state. A social group experiencing a chaotic period is vulnerable to a cascade of effects (Buzna *et.al*, 2007) from individual dyads. For example, rapid changes in dominance between individuals who interact over time could lead to instability in the entire group.

**Social Networks**

In the context of emergent complexity, there are two attributes of social networks that are critical to understanding the dynamic nature of social complexity. The first of these regards how information moves between members of a network, and the second involves the structural nature of these relationships themselves. Both of these attributes are affected by a number of network-specific mechanisms that govern the effects of social interaction and ultimately result in metastable network dynamics. This process is shown schematically in Figure 3.

**Spreading Activation.** Spreading activation models are often used to model the functioning of social networks over time, and have been anecdotally described as *six degrees of separation* (Watts, 2003). The basic functioning of a spreading activation model is simple; every network of conspecifics can be viewed as a series of nodes and connections. From a modeling perspective, each connection defines a communications channel, and responds to discrete interactions between two conspecifics over time. This can either be a unique response in time, or evolve trends of interaction. Connections may be broken off if they go unused for a certain period of time, while individuals with a high degree of connectedness may disproportionately establish connections with many new group members or individuals of a lower status over time (Newman, 2001).

Each conspecific will also have a threshold heuristic or selection criteria, which can be operationally defined as the ability to discriminate a stimulus or to recognize signals advertised by a conspecific. Among social insects, Seeley and Buhrman (2000) show how bees adaptively tune their waggle to draw attention of the other members of its hive to the location of resources. Ambiguity over whether or not certain stimulus is equivalent to or exceeds a given threshold can change either the threshold or the dyadic relationship.

Once certain nodes gain a disproportionate number of connections relative to the rest of the network, they take on the role of network hubs. These hubs control the flow of information and access to resources between subgroups such that the removal of a single hub can significantly affect the flow of information and the ability to establish weak ties. These effects are often called cascading effects (Barabasi, 2002), which can act to shift the system from one stable state to another (see Figure 3 for more).

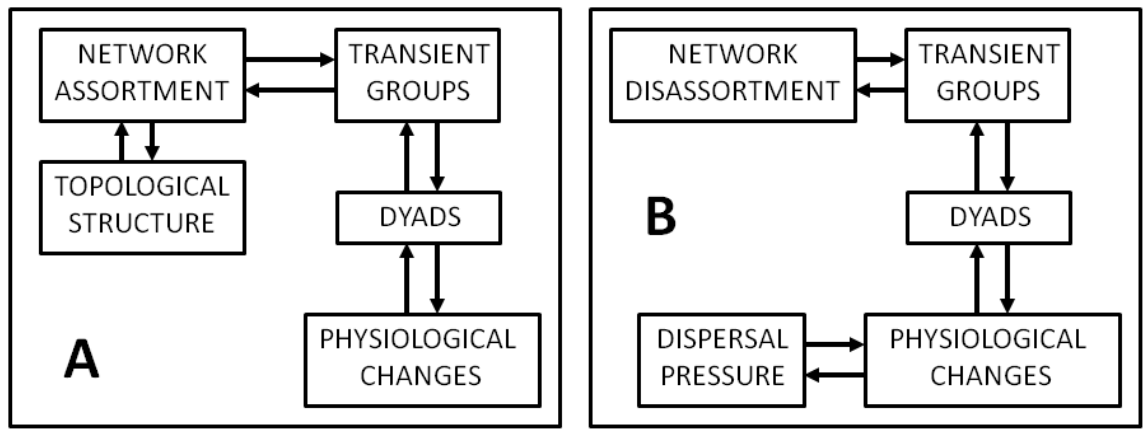

**Figure 3.** Two alternating mechanisms that results in the emergence of metastable network structures: network assortment (A) and network disassortment (B).

**Assortment and Weak Ties.** Newman (2003) proposes that certain types of dyadic interactions affect the stability and nature of the network as a whole. *Assortative mixing* (Figure 4) occurs when individuals with similar characteristics become more directly and exclusively associated with one another. This could include conspecifics with similar phenotypic markers or vocalizations. *Disassortative mixing* (Figure 4) occurs when dissimilar organisms become more directly associated with each other over time. Dyads use both public and private information to facilitate social learning and relationships in general, which contributes to the diffusion of information and information across a population. Most importantly, social relationships will develop over time and patterns of strong and weak ties will develop differentially throughout the network.

First-order connections are generally thought to be the most direct and permanent types of relationship in a social network (Newman, 2001, Watts, 2003), and stand in contrast to weak ties. One example from primates (Chapais, 2011) involves the difference between kin (strong first-order connections) and non-kin (weak first-order connections). However, Hauser and Marler (1993) point out that benefit based on the exchange of signals such as alarm calls and resources can be gained more easily by all members of a network and only require weak ties. More generally, Granovetter (1973) found disassortative relationships and weak ties (Csermely, 2006) to be important for maintaining sociality.

This is particularly important, as weak ties are potentially strong in that they allow access to both a variety of social subgroups and larger numbers of allies over time. Dissasortative mixing can be seen among male bonnet macaques, which were observed by Silk (1999) engaging in dyadic agonistic interactions and third-party alliances. More generally among primates, Dunbar (1992, 2001) argues that neocortical volume can be allometrically scaled to the dominant group size in a species. Related to what Silk (1999) has observed, social network size seems to be constrained by how many third-party relationships and coherent relationships a single individual can monitor. In addition, the stability of these relationships can be affected by hormonal (e.g. cortisol) concentrations (Sapolsky, 1992). Viewing this from a network perspective, Dunbar (2001) suggests that males serve as the weak links that hold female subgroups together which defined by stronger and more extensive first-order connections.

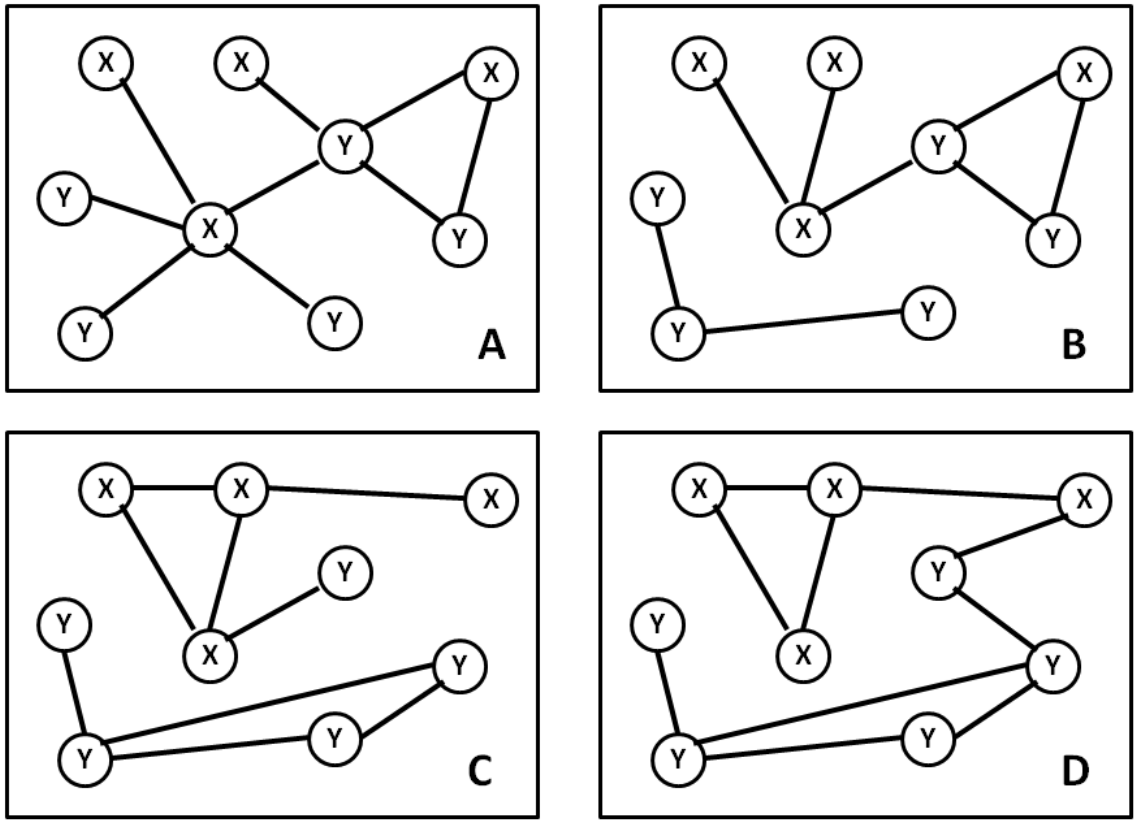

**Figure 4.** An example of assortative mixing in networks among two different classes of node (X and Y). Stepwise transition (A-D) from disassortative network topology (A) to assortative network topology with a weak tie (D).

## Collective Structures

In the next several sections, an assortment of swarming, flocking, and herding models are introduced and evaluated. All of these models essentially capture various elements of the processes involved in the self-organization of complexity social entities. Differences in each model reflect their differential application to specific animal groups, and are primarily based on different assumptions made about the cognitive and behavioral abilities of individual organisms. For example, while swarm intelligence models are indicative of social insects, they may of limited use in understanding mammalian societies. Likewise, models such as herding may require the organisms to possess a "theory of mind" level of intelligence, while a flocking model may not.

**Swarm Intelligence.** Swarm Intelligence refers to the collective action of animal social groups and the manner in which these interactions create something larger than the sum of their parts. Some of the literature in this area has treated computer simulations of algorithmic search functions and natural insect social complexity as overlapping constructs. As a result, swarming implies that groups of organisms can engage in cognitive tasks far more complex than what the individual can accomplish on its own (Kennedy and Eberhard, 2001). Therefore, applying such models outside the domain of social insects or using them to make generalizations about what is going on inside the organisms' head should be done with caution (Krause, Ruxton, and Krause, 2010).

Bonabeau *et al* (1997) have demonstrated through simulations and some empirical work that the buildup of insect social complexity through natural means can solve (e.g. converge upon a solution to) many complex problems. For example, Gambardella and Dorigo (1997) solved a *NP-hard* search problem in mathematics called the Traveling Salesman Problem (TSP) by using virtual ant pheromone trails. The problem can be posed as defining the shortest possible route between several spatial locations with the condition that each site can only be visited once. In the Gambardella and Dorigo (1997) model a population of ant agents is instructed to explore an exponentially large number of connections between all spatial locations randomly. To do so, pheromones were laid down on each trail evaluated so that its length was inversely related to the concentration of pheromone. After several thousand iterations of this process, an optimal route was found.

A similar process was found to contribute towards human trail formation by Helbing and Molnar (1997). It was assumed that initially, people randomly pass through a grass common to find a short cut to get to the other side. It was also assumed that simple heuristics are used to navigate across this open space, similar to the rule sets used in simulations of flocking birds and optimizing ants. This pseudo-random process of path integration resulted in the formation of visible paths over time (Helbing and Molnar, *19*97), which then acted as a feedback mechanism both by visually cuing future navigators Indirectly, this also lead to the maintenance and reinforcement of optimal paths over time. Thus, mental operations that make up cognition seem to be based on executing simple rules locally with help from objects in the environment and a vague sense of the global states for structures such as termite mounds (Bonabeau *et al*, 1998) or optimal paths.

These two examples are similar to what Seeley (1995) refers to as the "extended phenotype" of a social insect colony, and are shaped both by emergent complexity and cognitive forces. On the one hand, Hemelrijk (2000) has shown that even if the same set of rules are used by the same set of organisms, totally different structures representing a global state will result every time they interact. This suggests that historical forces also play a role in shaping the complexity seen at the social level.

However, it has also been shown in social insects that physical objects can mediate the interactions between organisms, ultimately leading to differential expressions of socially-relevant patterns in different contexts. This process of environment contributing to social complexity has been called stigmergy, and is common among social insects (Bonabeau, 1997, Camazine *et al, 19*98). Interestingly, stigmergy has also been used as an organizing principle in social robotics (Holland and Melhuish, 1999), and is consistent with the observation made by Clark (1997) that humans use a similar strategy he calls computational offloading to deal with complex interactions and tasks.

**Flocking.** This class of models most directly focuses on how all conspecifics in a group become engaged in coordinated behavior. Reynolds (1987) simulated flocking behaviors using the *boids* system by assigning each organism in a population three distinct behaviors. Collision avoidance was carried out by keeping a certain distance from all other neighboring flockmates. Velocity matching occurred when all individual organisms match the speed at which all neighboring flockmates are traveling. Finally, flock centering is an attempt by all organisms to stay close to each other by not straying too far from all neighboring flockmates. Each rule is applied in order of decreasing frequency as needed to maintain a coherent flock structure.

Each organism could also issue its own commands relative to the rest of the group. For example, acceleration requests such as "if I were in charge, I would move faster" were issued by different organisms simultaneously. If acceleration requests issued at a certain moment in time were contradictory for different organisms within the flock, they must cancel each other out to maintain the flocking structure. In the "boids" model, Reynolds (1987) uses a weighted average to reconcile these requests; an emergency situation might call for a queue of requests with the most pressing one satisfied first. In subsequent models based on a variety species (birds and insects) that use swarming during migration (Guttal and Couzin, 2010), other criterion are used. However, the local information used for establishing a swarm that exhibits network dynamics consists of very simple observations at the level of a dyad (Pais and Leonard, 2013).

It is important to remember that these structures are emergent and nonlinear, which means that the ubiquity of simple rules can produce order and regularity in social behavior (Hodgins and Brogan, 1994). For example, collision avoidance and velocity matching are complementary, and acts to maintain coordination among organisms. Flock centering refers to a center relative to a local group of organisms in space, which means that there are as many centers as there are organisms in the population. In general, all local groups are interrelated. This allows the flock to be flexible and highly modular, which may allow the flock to bifurcate and attain new stable states as it moves around barriers.

Interactions between individuals over space and time are driven by properties of the individual such as aggression, physiology, and memory (Chase *et al*, 1994) and observations of the environment. Thus, the actions of individual organisms or local groups in the flock can move it to a new stable state. Newman (2003) observes that humans in social networks act upon information gathered from first-order connections only. Partridge (1982) has found the same to be true for social networks within fish schools. This leads to hierarchical differentiation within the local population (Hemelrijk, 2000), which can either produce a cascade of effects or exist in a stable state. Rapid changes in dominance between individuals can destabilize the entire group.

According to Dunbar (1998), the "cognitive demands of social living" include regulating signaling behavior over time, especially with regard to vocalizations. Aubin and Jouventin (2000) study on Penguins showed that Penguin kin can recognize a distinct call embedded in a wall of noise. In this case, a large number of assorted and simultaneous penguin calls represent noise, and the "cocktail party" model of signal discrimination is used to measure how well individual penguins can identify conspecifics. In a similar fashion, Cynx and Gell (2004) have found that when random noise is introduced into their environment, zebra finches (*Taeniopygia guttata*) increase the amplitude of their vocalizations. Nightingales from the genus *Luscinia* also use amplitude regulation in the process of song learning (Brumm and Hultsch, 2001). Dunbar (1998) points out that vocal response to one's mate does not seem to simply be reflexive, but also modulated by the presence of conspecifics.

**Herding.** Herd behavior is defined by Banerjee (1992) as doing the same thing every other member of a social group is doing, even when private information is telling that individual something different. A theoretical example of this process involves a pool of individuals who get to choose between actions A, B, or C. Based on private information, which is the product of decision-making, there is a clear preference for executing B. However, all of the conspecifics who initially act upon their preferences demonstrate a minority preference for A.

Providing all other conspecifics have access to these public actions, the optimality of action B suddenly become ambiguous to the individual. Upon comparing private information with public information sampled from the outside world, A appears to be more optimal even though no change in the payoff for each action is immediately apparent. Thus, theory suggests that there is a defection from B or C to A by all subsequent individuals based on repeated adoption of what is being deployed in the real world. Thus, herding models may provide novel insights into the phenomenon of deceptive signaling.

Herd behavior, like flocking behavior is a positive feedback type system (Reynolds, 1987). That is, if a strategy A is somewhat successful or rewarding, then it is more likely to displace options B and C and create a homogeneous herd structure. However, if strategy A is immediately shown to be maladaptive, then subsequent adoption of A will be extremely limited. The concept of public information in animal social complexity is a bit different from its original application in human economic contexts, but still exists in various forms. According to Danchin *et al* (2004), information on foraging, breeding success, and mate-choice are all available to the general public in a social species. This can be done both by exploiting olfactory and sound cues placed in the environment and watching the outcomes of others' interactions in terms of eavesdropping on mating interactions or assessing predation risks. Emery and Clayton (2001) found that Scrub Jays possess a memory of location for food storages and a timing mechanism that informs them of when that resource can be best be taken from a conspecific. While this type of planned deception does not necessarily require formal neural mechanisms, it does require some internal representation of mental states.

**Game Theoretic Models**
Game theory can be defined as an evolutionary process and a mathematical method of analyzing interactive decision-making (Gintis, 2000). Game-theoretic relies most heavily on dyadic interactions, but can explain complexity at the individual and group levels as well (for a review of coevolutionary games, see Perc and Szolnoki, 2010). Theoretically speaking, two basic types of 'game' exist. Spatial games are where individuals interact in space, with the resulting topology defining the outcome. Temporal games allow for strategies to be deployed over time, yielding long-term trends and patterns. Games such as the Prisoners' dilemma (Axelrod, 1984) or Rock-Paper-Scissors (Sinervo and Lively, 1996, Maynard-Smith, 1996) fall into this category. As mentioned previously, the unit of analysis is generally a pair of organisms, although the strategy suite can be a product of both pure intentions by small groups or physiological changes within the organism. The following two examples show how game-theoretic models encapsulate social complexity in both space and time.

**Signaling and Spatial Game Theory.** Kraukauer and Pagel (1995) used spatial game theory to analyze the impact of limited mobility on the emergence of honest signaling. The authors assume that interactions between conspecifics are often in conflict, and on a more general level relates to the evolution of deception. If this is so, there are attempts by signalers to manipulate a receiver's preferences. In their paper on honest signaling, Kraukauer and Pagel (1995) used a 50 by 50 grid of empty cells was used to model four distinct strategies, consisting of two signaling strategies and two receiving strategies. Individuals can only communicate with their adjacent neighbors. There was heterogeneity in the population so that neighbors possess something of value to the neighbor. The individual advertised itself to the neighbor, and the neighbor had to accurately perceive that advertising level.

A receiver's fitness is maximized by minimizing the difference between the signaler's true state and its perceived state. The signaling strategies were discrete behaviors called honesty, dishonesty, gullibility, and suspicion. This model shows that honest signaling was favored even in the absence of any costs, and coexisted with dishonest strategies (Kraukauer and Pagel, 1995). A comparison of different paired strategy combinations over time showed that stable states evolve for each strategy over time.

In the Krakauer and Pagel (1995) and Reynolds (1987) studies, only the nearest neighbors had direct contact with the organism being studied. Some models such as that presented by Keitt and Johnson (1995) have solved this limitation through using mass action placement to represent the dynamic movement of organisms with differential behaviors over space and time. This is derived from the physics concept of diffusion-limited aggregation (Kauffman, 1993), and involves allowing artificial organisms called automata to diffuse across a grid.

The movement of these automata across the grid is governed by simple rules, and relies on spatial heterogeneity in terms of resources and placement of other automata over time to drive the formation of stable states and periods of systemic upheaval in time. Models that exhibit differential mortality and spontaneous segregation of automata types can also be used to understand predator-prey interactions over time (Keitt and Johnson, 1995). This model results in a qualitative signature on the grid surface that shows prey as a thin but dense wave front of automata, and predators as a diffuse trailing edge along one side of the "prey" wave front.

**Interactions modeled as a game.** In a more exclusively temporal application of game-theoretic models, Sinervo and Lively (1996) and Sinervo *et al* (2000) have modeled side-blotched lizard (*Uta stansburiana*) mating behaviors. Three morphs co-exist in the same population. Orange throated males are aggressive, defending large territories with many females. Blue throated males defend females aggressively, and hold smaller territories with fewer female mates overall. By contrast, yellow throated males patrol a large home range, and secretly copulate with females in the territories of dominant males. Throat color can signal relative status among conspecifics (Maynard-Smith, 1996). In this case, Sinervo *et al* (2000) showed that plasma testosterone plays a role in dominance. For example, orange throated morphs had naturally high levels of plasma T, and thus exhibited the "most dominant" strategy. Elevating levels of plasma T levels in blue and yellow morphs resulted in higher levels of endurance, activity, and home range size.

The dynamics of interactions between morphs can be captured using the non-transitive *rock-paper-scissors* game (Sinervo and Lively, 1996). In cases where phenotypic coloration is a mixed strategy, each polymorphism can be beneficial in different social situations. Each 'strategy' has a certain payoff, which can change over time as lizards adapted to a changing social environment. In this case, when there is an abundance of orange-throats, the incentive is higher for individuals to change strategies (and hence phenotypes). When phenotype (and hence strategy) is genetically determined, similar payoff dynamics lead to genetical rather than social selection. In the Sinervo *et al* (2000) study, each morph took on a pure strategy. Orange polymorphisms represent ultra-dominant behaviors, while yellow polymorphisms represent a sneaker strategy. In the case of side-blotched lizards, changes in the payoff over time give the game a coevolutionary quality. For example, if mostly orange morphs exist in population, then being orange is no longer advantageous and an increasing number of lizards will change into another throat color over time. This results in a series of evolutionary stable states (Maynard-Smith, 1982) across the species.

### Relational Sociality

Relational sociality can be defined as individual participation in dominance hierarchies and labor specialization. This can be either transient as was demonstrated in the cocktail-party model, or intimately connected to the development of neural structures. For example, O' Donnell *et al* (2004) studied the division of labor and development of a neural substrate in eusocial wasps (*Polybia aequatorialis*). Wasp tasks can be identified as working in the nest, moving to the nest periphery, and foraging. O' Donnell *et al* (2004) found that as in most social insects, tasks are associated with life-history developmental stages, and in particular with plasticity in mushroom bodies. Changes in mushroom bodies also play a role in sensory integration, learning, and memory.

As these structures develop, an organism gains the ability to perform different tasks which directly affects their social status. Another example of this is the findings of Edwards *et al* (2003) that body posture mediates social interactions in crayfishes produced by a balance of centralized interneurons and localized reflex systems. What is interesting about this example is that attack-and-approach behaviors reuse the neural circuitry for forward-and backward-walking, respectively. In honey bee mushroom bodies (Robinson *et al*, 1999), forager bees were found to possess larger calycal volume relative to Kenyon cell body size than when compared to nurse bees. This dimorphism, as in the case of eusocial wasps, is largely due to task performance. These examples ultimately beg the question of whether changes in neural structure are responsible for how organisms relate to one another, or if repeated sets of social interactions shape these structures.

To shed light on this dilemma, Edwards *et al* (2003) found that among crayfishes, small groups of six or fewer conspecifics tend to form linear dominance hierarchies. Among pairwise interactions within these hierarchies, dominance is defined mainly by relative body size. Large differences in body size generally lead to tail-flipping by the smaller organism. Tail-flipping also acts as an avoidance mechanism for smaller fish. By contrast, smaller differences in body size between two interacting fish leads to an escalating series of interactions of time. Agonistic interactions range from a defense posture to actual fighting, and act to further structure situations where ambiguous dominance relations exist (Goessmann *et al*, 2000).

In terms of the neural substrate, decisions to switch from aggressive to defensive behavior by a newly subordinate crayfish may reflect a shift from excitation to inhibition of pathways that control different behavior patterns. Long-term changes in social behavior may require longer-term neuromodulation. For example, Serotonin becomes inhibitory in subordinates but remains facilitatory in dominants after several weeks of pairing (Edwards *et al*, 2003). In long-term dominance hierarchies, submissives are observed to perform less tail-flipping but more backward-walking. Finally, Goessmann *et al* (2000) observed that in crayfish, long-term dominance relationships produce a lasting polarity in the outcome of agonistic bouts. Dyadic relationships can combine over time to form linear dominance hierarchies. In simulations produced by Hemelrijk (2000), stability in hierarchies over time is a consequence of feedback and interactions between spatial structure and the polarization of hierarchical positions.

**Emergent Feedback Sociality**

Feedback emergent sociality is similar to the process of Baldwinnian evolution (Deacon, 1998, Kerr, 2007). Baldwinnian evolution is based on interactions between the genotype, phenotype, behavior, and environment, and suggests feedbacks between environmental constraints, behavioral complexity, and the biological substrate. Feedback emergent sociality can be exemplified by examples from fishes and birds. Hoffman and Fernald (2000) showed that habitat instability triggers social interactions. This may cause changes in social status and affect reproduction in Lake Tanganyika cichlids. Jarvis *et al* (1998) looked at the social and genomic context of bird calls in zebra finches. ZENK gene expression in the anterior vocal pathway in the zebra finch brain has been found to vary according to the type of call a male zebra finch employed for different social purposes (Mello et al, 1992, Margoliash, 1997).

**Phenotypic Feedbacks and Social Status**

In the case of Hoffman *et al* (1999), the males have different phenotypes depending on the opportunity. In particular, this shows how behavioral interactions can modify cells in the brain. As one cichlid becomes dominant over another male, the dominant develops an eye stripe and bright yellow or blue coloration, controlled by cells like those in a chameleon. They also develop extra muscles and vertical black bars along the body, a dorsal fin tipped with red, a black spot on the tip of each gill cover and a reddish splotch in back of each gill. This dominant also shows about an eightfold increase in brain cells that produce gonadotropin-releasing hormone (GnRH), which governs sexual development. If another cichlid defeats the dominant male, the physiological changes reverse, and the previously territorial fish reverts to a female-like appearance (Hoffman and Fernald, 2000).

While these roles are generally conserved in vertebrates, specific behavioral effects vary with the social organization of individual groups. Looking at these types of sexual dimorphic and developmental mechanisms across vertebrates, Insel and Young (2000) found that the Oxytocin/Vasopressin family of peptides regulates sociosexual behaviors. For example, Vasotocin has been shown to increase vocalizations and aggression in territorial male sparrows, but does not contribute to similar increases in colony-dwelling zebra finches. Likewise, Vasopressin increases aggression and affiliation in mate guarding and monogamous prairie voles. However, this is not the case in montane voles, which do not guard mates and are promiscuous. In general, Vasopressin and Vasotocin are more abundant in males than in females. This sexual dimorphism is linked to a subset of neurons in the hypothalamus that are responsive to testosterone. Vasotocin cells also enlarge when sex-changing fish shift from a male to a female morph. Multiple morphs generally exhibit a correlation between the size of Vasotocin cells and male-like behavioral characteristics.

**Behavioral Genetics as Feedback**

Behavioral characteristics in social insects are also affected by a feedback between lower-level processes and environment. Krieger and Ross (2002) investigated the relationship between colony queen number and its genetic basis over time in social insects. Specifically, single-nucleotide variants of the gene *Gp-9* in fire ants codes for a pheromone-binding protein that plays a role in the chemical recognition of conspecifics. Krieger and Ross (2002) also suggest that when genetic polymorphisms coincide with different social forms, it may cause confusion among workers in terms of recognizing queens and regulating their numbers. As the sequence structure of *Gp-9* is generally conserved across South American fire ant species exhibiting different social types (Krieger and Ross, 2002), this finding suggests that even single genes can significantly interact with behaviors that lead to social complexity (see Robinson, Strambi, and Strambi, 1989).

Strong relationships between gene activity (measured using mRNA level) and social behavior has been observed in both bird and insect species. Patterns of singing-related gene activity (ZENK transcription factor) in several high-level brain areas of the adult zebra finch is dependent on whether a bird sings by itself or to another bird (Hessler and Doupe, 1999). This has implications as to the context of mating song, which can affect interactions a male bird can have with conspecifics. For example, ZENK expression has been found to be generally low when singing is female-directed. Female-directed activity is also accompanied by a courtship dance, which results in different patterns of electrophysiological activity. In social insects such as bees, the expression of *per* can determine the social structure (e.g. division of labor) observed in colonies and other collective groups (Toma, Moore, Bloch, and Robinson, 2000). In the case of ants, zebra finches, and bees, the genome plays an integral role in structuring social complexity and related interactions.

Social context also influences singing-related electrophysiological activity. In turn, electrophysiological activity can modulate subsequent gene expression. In the anterior forebrain nuclei L-MAN and Area X, gene activity during directed singing is lower in magnitude and more consistent in pattern across renditions than activity during undirected song. Strong modulation of neural activity in HVc and RA would be likely to result in changed song output. The lack of a major difference in motif structure between directed and undirected singing raises the possibility that social modulation of neural activity is weaker in these motor nuclei than in the anterior forebrain. Further evidence of the link between organismal and social complexity has been demonstrated by Karten (1991), who has shown that neural substrate associated with the zebra finch anterior vocal pathway is homologous with the mammalian neocortex and basal ganglia in mammals.

**Conclusions**
Each proposed signature should be thought of in relation to a larger theoretical framework. In simulations and observations of actual animal groups alike, a distinction can be made between specifications for individuals in a set of organisms and behavior that results as organisms interact with one another in the context of a specific environment (Langton, 1989). It is also important to realize that models of emergent complexity requires of a minimum of two analytical scales. The first should describe the conditions of complexity and constraints on system dynamics, while the second should describe global behavior. These levels of analysis can be clarified and better understood by using robotics models (Garnier, 2011) in tandem with computational models of the physiological substrate. Such models might also clarify the role of gene expression and genomic structure with regard to social behaviors (Robinson, 2002).

Models of social complexity are even powerful when they account for interactions between both conspecifics and other social groups simultaneously. Taking into account concepts such as group selection, selective exchanges of conspecifics between social subgroups, and social accounting mechanisms (Dunbar, 1992) will lead to more robust models in general. In general, computational representations of behavior must be well-specified (Goldstone and Janssen, 2005). Otherwise, modeling emergent complexity becomes a *garbage in, garbage out* enterprise. Good models should ultimately yield good quantitative and qualitative signatures, whether they consist of realistic looking herding and flocking or highly adaptive swarming.

One outstanding issue realted to this model is the relationship between social network topological complexity and lower-level mechanisms and outcomes. For example, network topologies are initially shaped by local dyadic interactions that determine (at least in part) the degree of global assortative interaction. However, network topologies are also continually updated by information flow. As this theory predicts, most of this information comes from the physiological states and background inherent in the interacting conspecifics. Yet collective behavior is also constrained by the global network topology. The role of a conspecific's position in the network topology (e.g. weak tie *vs.* network hub) might act as a closed-loop feedback on their physiological state and even the nature of transient social structures. Depending on the critical nature of this role with regard to changing or maintaining the social milieu (Buchanan, 2002), such individuals (and their physiological signature) may play an outsized role in the self-organization and regulation of sociality.

Applying this theory to natural systems has numerous advantages and drawbacks. One potential caveat for viewing social complexity in this way is that certain principles and models may not apply evenly across the phylogeny of animal diversity or even in different contexts within the same taxonomic group. The examples presented above do not focus on any one group, but also does not explicitly assume a certain level of intelligence beneath which these principles of self-organization do not apply. While it could be argued that certain interactions require a certain level of cognitive complexity, it has been shown here to some extent that behaviors leading to complexity can arise in novel ways, perhaps even calling into question the link between social complexity and cognitive complexity.

*guttata*. *Animal Behaviour,* 67, 451-455 (2004).

Csermely, P. Weak links: stabilizers of complex systems from proteins to social networks. Springer Verlag, Berlin (2006).

Danchin, E., Giraldeau, L-A., Valone, T.J., and Wagner, R.H. Public Information: from nosy neighbors to cultural evolution. *Science,* 305, 487-491 (2004).

Deacon, T.W. The Symbolic Species: the coevolution of language and the brain. W.W. Norton, New York (1988).

de Vries, H. and Biesmeijer, J.C. Self-organization in collective honeybee foraging: emergence of symmetry breaking, cross inhibition and equal harvest-rate distribution. *Behavioral Ecology and Sociobiology,* 51(6), 557-569 (2002).

Dunbar, R. I. M. Neocortex size as a constraint on group size in primates. *Journal of Human Evolution*, 20, 469-493 (1992).

Dunbar, R.I.M. The social brain hypothesis. *Evolutionary Anthropology,* 6, 178-190 (1998).

Dunbar, R.I.M. Neocortex size and social network size in primates. *Animal Behaviour,* 62, 711-722 (2001).

Edwards, D.H., Issa, F.A., and Herberholz, J. The Neural Basis of Dominance Hierarchy Formation in Crayfish. *Microscopy Research and Technique,* 60, 369–376 (2003).

Emery, N.J. and Clayton, N.S. Effects of experience and social context on prospective caching strategies by scrub jays. *Nature,* 414(6862), 443-445 (2001).

Gambardella, L.M. and Dorigo, M. Ant-Q: a reinforcement learning approach to the Traveling Salesman Problem. IN Proceedings Twelfth International Conference on Machine Learning, pp. 252-260. Volume ML-95. Morgan Kaufmann, Palo Alto, CA (1995).

Garnier, S. From Ants to Robots and Back : How Robotics Can Contribute to the Study of Collective Animal Behavior. IN Bio-inspired Self-organizing Robotic Systems, Studies in Computational Intelligence, 355, pp. 105-120 (2011).

Gintis, H. Game Theory Evolving: a problem-centered introduction to modeling strategic interaction. Princeton University Press, Princeton, NJ (2000).

Goessmann, C., Hemelrijk, C., and Huber, R. The formation and maintenance of crayfish hierarchies: behavioral and self-structuring properties. *Behavioral Ecology and Sociobiology,* 48, 418-428 (2000).
18